\documentclass[
  journal=jacsat,    
  manuscript=article,
  articletitle=true
]{achemso}
\usepackage[utf8]{inputenc}
\usepackage[T1]{fontenc}
\usepackage{amsmath,amssymb,amsfonts}
\usepackage{graphicx}
\usepackage{dcolumn}
\usepackage{bm}
\usepackage[colorlinks=true,allcolors=blue]{hyperref}
\usepackage{physics}
\usepackage{xcolor}
\usepackage{float}
\usepackage{siunitx}
\usepackage{tikz}
\usepackage{mathtools}
\usepackage[version=3]{mhchem} 
\usepackage{subfigure} 
\usepackage{float}

\author{Eric~R.~Bittner}
\email{ebittner@central.uh.edu}
\affiliation[UH]{Department of Physics, University of Houston, Houston, TX~77204, United~States}
\alsoaffiliation[IC-UdeM]
{Institut Courtois, Universit\'e de Montr\'eal, 1375 Avenue Th\'er\`{e}se-Lavoie-Roux, Montr\'eal H2V~0B3, Qu\'ebec, Canada}

\author{Carlos~Silva-Acu\~{n}a}
\email{carlos.silva@umontreal.ca}
\affiliation[IC-UdeM]
{Institut Courtois, Universit\'e de Montr\'eal, 1375 Avenue Th\'er\`{e}se-Lavoie-Roux, Montr\'eal H2V~0B3, Qu\'ebec, Canada}

\title{Coherent Spectroscopic Probes of Topology: A Velocity-Gauge Perspective}

\begin{tocentry}
\centering
\includegraphics[width=9cm,height=4.5cm,keepaspectratio]{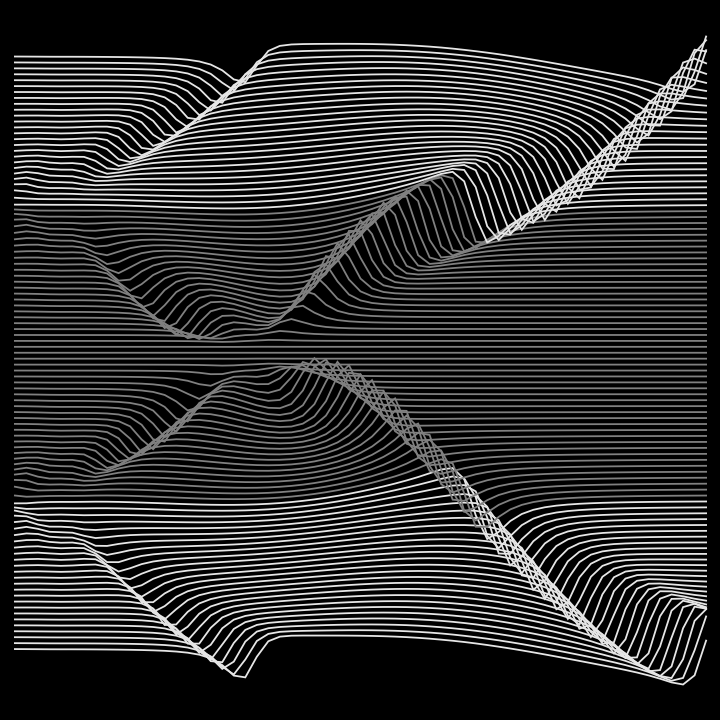}
\end{tocentry}

\begin{document}

\begin{abstract}
We present a velocity-gauge formalism for computing nonlinear current response functions in periodic systems and apply it to the Su–Schrieffer–Heeger (SSH) model as a minimal topological testbed. By retaining the full minimal coupling Hamiltonian and avoiding the rotating wave approximation, we construct gauge-consistent expressions for the linear and third-order current susceptibilities using retarded Green's functions. Our results reveal how nonlinear optical spectra encode not only energy-level transitions but also interband phase coherence and topological winding. In the topological phase, the third-order response exhibits characteristic phase inversions and spectral asymmetries that are absent in the trivial phase. These features reflect geometric changes in the Bloch eigenstates and highlight the role of virtual pathways in shaping the nonlinear signal. Our framework offers a robust and extensible platform for modeling nonlinear light–matter interactions in topological 
materials beyond the dipole approximation and the standard Coulomb-gauge formulation in molecular spectroscopy.
\end{abstract}

\section{Introduction}

Coherent optical spectroscopy, specifically multidimensional spectroscopy, is a powerful tool for interrogating ultrafast dynamics and correlated electronic states in molecules and materials~\cite{Mukamel1995,li2023optical,fresch2023two}. 
Molecular~\cite{kelley2022condensed} and semiconductor~\cite{klingshirn2012semiconductor} spectroscopy is generally formulated in the Coulomb gauge, because it permits intuitive separation of longitudinal and transverse fields that produce local action of light, as the vector potential $\vec{A}(\vec{r},t)$ naturally describes transverse (radiative) electromagnetic fields, while the scalar potential $\phi$ governs instantaneous electrostatic interactions. This is advantageous in that the light-matter interactions can be treated perturbatively with respect to the vector potential, leading to the dipole approximation, 
and in the interaction picture, quantum equations of motion for photoexcitations are developed with respect to the transition dipole operator~\cite{hamm2011concepts}. This picture breaks down in the case of the ultrastrong light-matter coupling regime~\cite{frisk2019ultrastrong,forn2019ultrastrong,le2020theoretical,stokes2021ultrastrong}, in which the system-light interaction Hamiltonian cannot be expanded perturbatively, with impunity, with respect to the vector potential.
In an alternative formulation it is often convenient to implement the length gauge~\cite{Cohen_Tannoudji_atomphoton}, in which the electric field couples to the electric field, as opposed to the need to directly invoke the vector potential that couples to the canonical momentum as in the Coulomb gauge. This naturally permits expansion into multipole components beyond the dipole approximation. This is the preferred treatment in the ultrastrong light-matter coupling regime and other strongly dipole-dominated contexts.

Topological materials are a fascinating class of quantum systems whose defining features are not set by local order --- such as atomic arrangements or symmetry breaking --- but by global, gauge-invariant properties of their electronic wavefunctions~\cite{tanaka2015short,wang2017topological}. This is of high relevance not only in quantum condensed matter, but also in chemistry~\cite{popelier2016quantum,kumar2020topological,cano2021band}. Indeed, a helpful analogy comes from chemistry: consider polyacetylene, a linear chain of carbon atoms with alternating single and double bonds. This system can exist in two distinct bonding patterns (dimerizations), which are not smoothly connected to each other --- they represent topologically distinct ground states. At the interface between these two phases, a localized electronic state emerges. This boundary state is robust: it does not disappear unless the entire electronic structure is qualitatively changed. This is a molecular-scale realization of a topological defect. More generally, in crystalline solids, as an electron evolves around a closed loop in momentum space, its wavefunction can accumulate a geometric (Berry) phase, much like phase shifts in cyclic molecular reactions or aromatic ring currents. These phases are not arbitrary --- they reflect the topological structure of the band. Topology thus classifies materials not just by symmetry, but by quantized global invariants, like winding numbers or Chern numbers, which are preserved unless the band gap closes. This leads to remarkable physical consequences: protected boundary states that conduct electricity at the edges while the bulk remains insulating, quantized transport coefficients, and responses to light or fields that are unusually geometric and robust --- immune to disorder and thermal noise. Importantly, topological features are not restricted to exotic materials. They can emerge in tight-binding models, molecular crystals, or engineered nanostructures built from organic components or low-dimensional materials. 
These topological features are now systematically classified via the formalism of topological quantum chemistry, which connects real-space orbital content to symmetry-protected band structures in crystalline solids~\cite{Bradlyn2017}.

For physical chemists, this opens an exciting frontier where ultrafast and nonlinear spectroscopy can probe not just electronic structure, but the topological character of materials, with potential applications in molecular electronics, quantum information, and beyond. 
Topological materials exemplify how global properties and collective behaviors emerge from diverse local interactions. In these systems, the overall electronic properties are not dictated by individual atomic sites but by the holistic arrangement and interactions across the material. Just as topological invariants confer robustness to materials against local perturbations, fostering an inclusive and equitable environment strengthens a community's resilience and adaptability. By embracing diverse contributions, both in materials science and societal structures, we can achieve systems that are not only robust but also rich in functionality and innovation.

In this Perspective, we outline the foundations and practical implementation of the velocity-gauge formalism for computing linear and nonlinear 
responses in molecular and condensed-phase systems. We emphasize its utility for analyzing light–matter coupling in strongly dipole-allowed systems beyond the dipole approximation, and demonstrate its application to prototypical models such as the Su–Schrieffer–Heeger (SSH) chain--which exhibits both trivial and topological phases. Our goal is to bridge the gap between advanced gauge-consistent theories and spectroscopic observables relevant to chemical physics in which specific signatures of topology are relevant.  
Following a brief overview of the role of gauge in 
electrodynamics and light-matter coupling, we will work in the Coulomb gauge and explicitly evaluate optical response functions from current–current correlations in the velocity gauge framework.

\subsection*{Gauge Invariance and Light--Matter Interactions: A Tutorial Over\-view}

Understanding the interaction between light and matter in quantum systems requires careful attention to the role of electromagnetic gauge fields and their representation in the Hamiltonian. The standard approach begins with the principle of minimal coupling, in which the canonical momentum operator \( \vec{p} \) is replaced by \( \vec{p} - q\vec{A}(\vec{r},t) \), where \( \vec{A} \) is the vector potential and \( q \) is the particle charge. This yields the time-dependent Hamiltonian:
\begin{equation}
H = \frac{1}{2m} \left[ \vec{p} - q\vec{A}(\vec{r},t) \right]^2 + q\phi(\vec{r},t) + V(\vec{r})
\end{equation}
where \( \phi(\vec{r},t) \) is the scalar potential, and \( V(\vec{r}) \) represents any additional binding or lattice potential.
Gauge invariance asserts that physical observables are unaffected by local transformations of the potentials:
\begin{align}
\vec{A}' &= \vec{A} + \nabla \Lambda(\vec{r},t) \\
\phi' &= \phi - \partial_t \Lambda(\vec{r},t)
\end{align}
for any scalar gauge function \( \Lambda(\vec{r},t) \). Different choices of \( \Lambda \) define different gauges.
The choice of gauge affects how one formulates perturbation theory and computes observable quantities, but physical predictions must remain invariant if all orders are treated consistently. In practice, numerical implementations often truncate the Hilbert space or neglect diamagnetic terms, introducing gauge-dependent discrepancies.
Two commonly used gauges in quantum optics and solid-state physics are
the Coulomb (or velocity) gauge 
in which we set $\nabla\cdot A = 0$, with $\phi=0$
for purely transverse fields,
and the Length 
gauge in which we set $\Lambda(\vec r,t) = -\vec{r}\cdot \vec{A}$, 
which eliminates the vector potential 
and the external 
electric field enters via the 
scaler potential. Table~\ref{tab:1} gives a brief comparison of the two gauges.

\begin{table}[ht]
\begin{center}
\begin{tabular}{|l|l|l|}
\hline
\textbf{Feature} & \textbf{Length Gauge} & \textbf{Velocity Gauge} \\
\hline
Interaction term & $-q \vec{r} \cdot \vec{E}(t)$ & $-\frac{q}{m} \vec{A}(t) \cdot \vec{p}$ \\
Gauge condition & $\vec{A} = 0$, $\phi \neq 0$ & $\phi = 0$, $\vec{A} \neq 0$ \\
Matrix elements & $\langle f | \vec{r} | i \rangle$ & $\langle f | \vec{p} | i \rangle$ \\
Applicability &  bound states &  continuum states \\
Spectral convergence & Low-frequency regime & High-frequency/strong field regime \\
\hline
\end{tabular}
\end{center}
    \caption{Comparison of the length and velocity
    gauges}
    \label{tab:1}
\end{table}

 In the length gauge, the electric field couples directly to the dipole moment of the system via the interaction Hamiltonian $H_{\text{int}} = -q \vec{r} \cdot \vec{E}(t)$. This form is intuitive and computationally convenient in systems with well-localized electronic orbitals, particularly molecules where the characteristic 
 length-scale of the system (e.g.\ 1\,nm) 
 is small compared to the wavelength of the perturbing 
 radiation field (e.g.\ 100--1000\,nm). This would naturally extend to other systems with strong local couplings, such as strongly correlated materials. 
While this framework has proven highly successful for describing multiphoton processes in atomic, molecular, and small condensed-phase systems under weak to moderate excitation conditions, the position operator $\vec{r}$ is ill-defined in periodic systems, leading to inconsistencies in the evaluation of optical response in solids or extended media.
Consequently, the dipole approximation begins to break down in several regimes now routinely accessed by modern spectroscopic techniques: (i) when electronic wavefunctions extend over length scales comparable to the driving optical wavelength, as in nanostructures or Rydberg states; (ii) in the presence of intense, tightly focused pulses where spatial field gradients become significant; (iii) when employing structured or high-frequency light sources—such as attosecond pulses, vortex beams, or XUV radiation—whose spatial variation can no longer be neglected.

In the Coulomb or velocity gauge, the light–matter interaction enters
when the canonical momentum $\vec{p}$ is shifted by the vector potential $\vec{A}(\vec{r},t)$:
\begin{equation}
H = \frac{1}{2m} \left[ \vec{p} - q\vec{A}(\vec{r},t) \right]^2 + V(\vec{r}).
\end{equation}
This yields both paramagnetic ($\vec{p} \cdot \vec{A}$) and diamagnetic ($\vec{A}^2$) terms, which play essential roles in governing nonlinear responses and preserving gauge invariance.
Within this framework, physical observables—such as high-harmonic spectra, photoemission currents, and transient absorption—are derived from current–current correlation functions rather than polarization-based susceptibilities.
It is also a natural framework for extended systems with 
band structure and periodicity. 
Expanding the square yields both the paramagnetic term ($\vec{p} \cdot \vec{A}$) and the diamagnetic term ($\vec{A}^2$):
\begin{align}
H = H_0 + H'(t) = \frac{\vec{p}^{,2}}{2m} + V(\vec{r}) - \frac{q}{m} \vec{p} \cdot \vec{A}(t) + \frac{q^2}{2m} \vec{A}^2(t).
\end{align}
The paramagnetic term $-\vec{p} \cdot \vec{A}$ gives rise to odd-order nonlinearities and defines the current operator, while the diamagnetic term $\vec{A}^2$ becomes significant for intense or broadband fields.  
The term  becomes important at high intensities and for long wavelengths, and can no longer be neglected without introducing inconsistencies.
To preserve gauge invariance, it is essential to retain the full structure of $H'(t)$ when computing response functions, especially in the presence of strong fields, delocalized wavefunctions, or topological phases. 
It is also 
important to 
point out that
the Coulomb
(or transverse) gauge ensures 
that operator ordering is symmetric: $\nabla\cdot \vec{A} = 0 \Rightarrow [p_\mu,A_\mu] =0$.
 We can see this by considering the action 
 of $(\vec{p} \cdot \vec{A} + \vec{A} \cdot \vec{p})$ on the quantum wavefunction. First,  $\vec{p}\cdot\vec A \psi = -i\hbar\vec A\cdot \nabla\psi$ 
 and that 
 $\vec{A} \cdot \vec{p} \psi = -i\hbar \left( \vec{A} \cdot \nabla \psi + (\nabla \cdot \vec{A}) \psi \right)$.
Setting
 $\nabla\cdot \vec{A}=0$ 
 (as per the Coulomb gauge) gives the symmetric relation.

A rigorous velocity-gauge treatment~\cite{Reiss2000StrongFieldGauge, Chelkowski2005HHGGauge} is required to compute the High Harmonic Generation (HHG) spectrum accurately and to ensure gauge invariance of observable quantities. Notably, non-dipole effects in HHG have been experimentally observed in gases and solids, especially when driven with mid-IR pulses or focused beams with tight spatial profiles~\cite{Krausz2009AttosecondScience, Hofmann2019NonDipoleHHG}.
Recent experimental platforms—including time-resolved ARPES, high-harmonic generation in solids, and multidimensional coherent spectroscopy with structured light—have revealed clear signatures of dipole breakdown, prompting the need for a consistent velocity-gauge formalism in quantum chemical modeling. Notably, recent theoretical work~\cite{Liu2020trARPESGauge, DeGiovannini2016VelocityGaugeARPES} has demonstrated that only velocity-gauge calculations can reproduce the experimentally observed momentum shifts and emission asymmetries in strong-field tr-ARPES, particularly in layered or topologically nontrivial materials where surface states couple strongly to the in-plane components of $\vec{A}(\vec{r},t)$. These developments motivate a broader adoption of current-based response theories in the physical chemistry and spectroscopy communities.

\section{Theory: Minimal Coupling and Gauge Structure}

Having discussed the Coulomb gauge, we consider a general quantum system coupled to a classical electromagnetic field where $\nabla \cdot \vec{A} = 0$ and the scalar potential $\phi$ is taken to vanish. 
The observable response to the applied fields is 
the current as opposed to the macroscopic polarization 
used in usual treatments of non-linear spectroscopy using the length-gauge. Here, 
the expectation value of the current is defined in terms of the paramagnetic current operator,
\begin{align}
\hat{\jmath}_\mu(t) = -\frac{q}{m}  p_\mu(t).
\end{align}
This can be expressed as a perturbative expansion in powers of the vector potential:
\begin{align}
\langle \hat{\jmath}_\mu(t) \rangle &= \int_{-\infty}^{\infty} dt_1 , \chi^{(1)}_{\mu\nu}(t,t_1) A_\nu(t_1) \\
&+ \iiint_{-\infty}^{\infty} dt_1 dt_2 dt_3 \chi^{(3)}_{\mu\nu\rho\sigma}(t, t_1, t_2, t_3)
A_\nu(t_1) A_\rho(t_2) A_\sigma(t_3) + \cdots
\end{align}
where $\chi^{(1)}_{\mu\nu}$ and $\chi^{(3)}_{\mu\nu\rho\sigma}$ are the linear and third-order current response functions.
Note, that we neglect the $\chi^{(2)}$ 
contribution.  This term vanishes for all
systems with a center of inversion.
Many two-dimensional systems, such as \ce{MoS2} and \ce{WS2} 
lack inversion symmetry and consequently 
have large $\chi^{(2)}$ responses, as do
various topological insulators, Janus monolayers, 
and ferroelectric 2D materials~\cite{AversaSipe1995}.

\subsection{Linear responses}

The expectation value of the induced current in response to an external vector potential \( \vec{A}(\vec{r}, t) \) is given by
\begin{equation}
\langle \hat{\jmath}_\mu(\vec{r}, t) \rangle = \int d^3 r' \int dt' \, \chi^{(1)}_{\mu \nu}(\vec{r}, t; \vec{r}', t') A_\nu(\vec{r}', t') + \cdots,
\end{equation}
where the linear response function is defined via the Kubo formula,
\begin{equation}
\chi^{(1)}_{\mu \nu}(\vec{r}, t; \vec{r}', t') = -\frac{i}{\hbar} \theta(t - t') \left\langle \left[ \hat{\jmath}_\mu(\vec{r}, t), \hat{\jmath}_\nu(\vec{r}', t') \right] \right\rangle_0.
\end{equation}
For a homogeneous system, translational symmetry permits a Fourier representation,
\begin{equation}
\chi^{(1)}_{\mu \nu}(\vec{q}, \omega) = \int d^3 r \int dt \, e^{i(\omega t - \vec{q} \cdot \vec{r})} \chi^{(1)}_{\mu \nu}(\vec{r}, t; \vec{0}, 0),
\end{equation}
and the linear current response simplifies to
\begin{equation}
\hat{\jmath}_\mu(\omega) = \frac{q^2}{m} \chi^{(1)}_{\mu\nu}(\omega) A_\nu(\omega),
\end{equation}
where \( \chi^{(1)}_{\mu\nu}(\omega) \) denotes the current susceptibility in the velocity gauge.

In the length gauge, the response is given by the polarization \( P_\mu(\omega) \) induced by the electric field \( E_\nu(\omega) \),
\begin{equation}
P_\mu(\omega) = \chi^{(1)}_{\mathrm{P}, \mu \nu}(\omega) E_\nu(\omega).
\end{equation}
The current and polarization susceptibilities are related by
\begin{equation}
\chi^{(1)}_{\mu \nu}(\omega) = i\omega \chi^{(1)}_{\mathrm{P}, \mu \nu}(\omega),
\end{equation}
as \( j(\omega) = -i\omega P(\omega) \).

In principle, the position operator $\vec{r}$ and the momentum operator $\vec{p} \sim -i\hbar \nabla$ have the same transformation properties under spatial inversion, and thus yield equivalent selection rules when used consistently in a complete basis. However, practical differences emerge in truncated Hilbert spaces or periodic systems where $\vec{r}$ is ill-defined. In such cases, the velocity gauge---formulated in terms of the current operator $\vec{j}$ or the momentum matrix elements---avoids ambiguities and ensures manifest gauge invariance. Moreover, when higher-order terms such as $\vec{A}^2$ and field gradients are retained in the minimal-coupling Hamiltonian, velocity-gauge formulations naturally capture non-dipole effects including even-order harmonic generation and spatially dependent phase accumulation, which are often inaccessible or neglected in standard dipole-based models.

\subsection{Non-linear responses}
We now consider the non-linear responses. 
Switching to the delay variables with \( t_3 = 0 \), and defining \( \tau_1 = t_2 - t_3 \), \( \tau_2 = t_1 - t_2 \), and \( \tau_3 = t - t_1 \), we write:
\begin{align}
\hat{\jmath}_\mu(t) = \int_0^\infty d\tau_1 \int_0^\infty d\tau_2 \int_0^\infty d\tau_3 \,
\chi^{(3)}_{\mu \nu_1 \nu_2 \nu_3}(\tau_3, \tau_2, \tau_1) \,
A_{\nu_1}(\tau_1) A_{\nu_2}(\tau_1 + \tau_2) A_{\nu_3}(0).
\end{align}
where $\tau_1$, $\tau_2$, and $\tau_3$ are the time-intervals between the interactions.
The structure of the third-order current response in the velocity gauge closely parallels the well-established polarization response formalism used in nonlinear optics. Both are built on nested commutators of the system observable with time-evolved interaction operators and naturally organize into time-ordered integrals over causal delays. 
Consequently, the third-order susceptibility in the velocity gauge, with frequency-domain treatment of the first and third time intervals and time-domain treatment of the central delay \( \tau_2 \), can be written as:
\begin{align}
\chi^{(3)}_{\mu \nu_1 \nu_2 \nu_3}(\omega_3, \tau_2, \omega_1) =
\left( \frac{i}{\hbar} \right)^3
\operatorname{Tr} \left\{
\hat{\jmath}_\mu 
\left[\hat{\jmath}_{\nu_1},
\tilde G^{\text{ret}}(\omega_3) 
\left[ \hat{\jmath}_{\nu_2}(\tau_2), \,
\tilde G^{\text{ret}}(\omega_1) \left[ \hat{\jmath}_{\nu_3}, \rho_{\text{eq}} \right] \right]\right]
\right\},
\end{align}
where
\( \tilde G^{\text{ret}}(\omega) = (\omega - \mathcal{L}_0 + i \eta)^{-1} \) is the retarded Liouvillian Green’s function, 
\( \hat{\jmath}_{\nu_i}(\tau) = e^{i \mathcal{L}_0 \tau} \hat{\jmath}_{\nu_i} \) 
is the Heisenberg-evolved current operator,
\( \mathcal{L}_0 \rho = [H_0, \rho] \) is the unperturbed Liouvillian,
and 
\( \rho_{\text{eq}} \) is the thermal equilibrium density matrix. The full response is given by 
all of the nested commutator terms and
represents a fully time-ordered sequence of field-matter interactions. Each commutator corresponds to a field-induced transition, while the nested retarded Green's functions propagate the density matrix forward in time between events.
This structure directly maps onto the diagrammatic expansion in terms of double-sided Feynman diagrams, where each  
$\hat \jmath_{\nu}$
denotes an interaction vertex and each $\tilde G^{\text{ret}}$ represents temporal evolution on the ket or bra side. Different permutations of the interaction sequence correspond to different Liouville pathways such as ground-state bleaching (GSB), stimulated emission (SE), and excited-state absorption (ESA).

Despite the asymmetric appearance of the commutator structure, the full third-order susceptibility obeys well-defined symmetries under permutations of field indices and time orderings. For example, rephasing and non-rephasing pathways can be obtained by reorganizing the time arguments $(\omega_1,\tau_2,\omega_3)$, and the susceptibility satisfies Kramers--Kronig relations and complex conjugation symmetry. Exchange of $\nu_1\leftrightarrow \nu_3$ is related to time-reversal conjugate diagrams.
Furthermore, the tensor structure of $\chi^{(3)}$ is constrained by the point-group symmetry of the system, as well as the gauge and field configurations. 
We next apply this approach to a non-trivial
two-band system that can be treated analytically.

\subsection{Model Calculations: SSH Model}

The Su–Schrieffer–Heeger (SSH) model describes a one-dimensional dimerized lattice with alternating hopping amplitudes between neighboring sites~\cite{Su1979}. Originally introduced to explain solitons in polyacetylene, the SSH model has become a paradigmatic system for studying topological phases of matter. Its two distinct phases—trivial and topological—are distinguished by a quantized Zak phase and the presence or absence of zero-energy edge states under open boundary conditions~\cite{Zak1989}. The SSH model has found broad application in contexts ranging from charge fractionalization in polymers to photonic crystals~\cite{Lu2014}, cold atom systems~\cite{Meier2016}, and topoelectronic devices~\cite{Barcikowski2021}. Its simplicity and analytical tractability make it an ideal platform for exploring band topology, symmetry-protected edge modes, and the interplay between geometry and electronic response in low-dimensional systems.
For our purposes here, it offers a non-trivial model for highlighting how a velocity gauge 
can be sensitive to topological information 
of the system. 
Within the SSH model, interband matrix elements 
\( \langle u_c(k) | j(k) | u_v(k) \rangle \) depend on the winding of the pseudospin vector \( \hat{d}(k) \), reflecting the topological character of the system. As the model transitions from trivial to topological, these matrix elements acquire geometric phase shifts that modulate both amplitude and phase of the nonlinear pathways. Consequently, features such as line shape asymmetry, peak intensity shifts, and phase inversions emerge across the topological transition.

\begin{figure}[th]
    \centering
\begin{tikzpicture}
\node[anchor=north west] at (0,0) {\includegraphics[width=6cm]{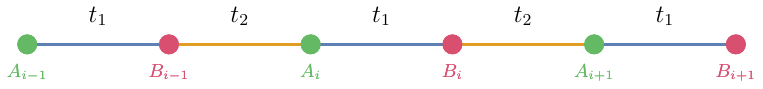}};
\node[anchor=north west] at (0,-1.) {\includegraphics[width=6cm]{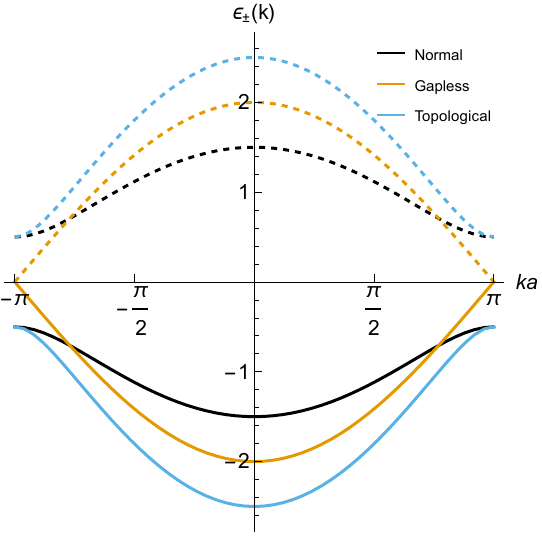}};
\node at (-0.5,0.2) {\textbf{(a)}};

\node[anchor=north west] at (6.5,-1.27) {\includegraphics[width=6cm]{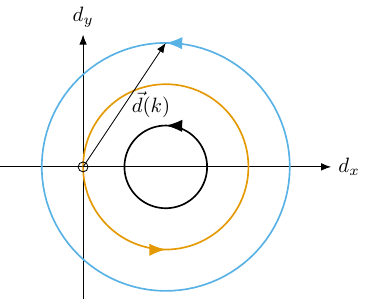}};
\node at (7.0,0.2) {\textbf{(b)}};
\end{tikzpicture}
    \caption{Sketch of the SSH Model. (a)The 
    top shows the alternating lattice sites  $A$, and $B$ and the respective hopping integrals $t_1$ and $t_2$.  The bottom shows the
    valence and conduction bands for the 
    normal ($t_1>t_2$), gapless ($t_1=t_2$),
    and topological ($t_1<t_2$) regimes. 
    (b) Hamiltonian trajectory of the pseudospin vector, $\vec{d}(k)$,
    for the normal/trivial, gapless, and topological regimes.
    Colors correspond to scheme in (a), arrows indicate
    the direction of increasing $k$.}
    \label{fig:SSHModel}
\end{figure}

In real space, the SSH Hamiltonian reads:
\begin{align}
H = \sum_n \left[ t_1\, c^\dagger_{A,n} c_{B,n} + t_2\, c^\dagger_{A,n+1} c_{B,n} + \text{h.c.} \right],
\end{align}
where \( c^\dagger_{A,n} \) and \( c^\dagger_{B,n} \) create electrons on sublattices \( A \) and \( B \) in unit cell \( n \), and \( t_1 \), \( t_2 \) are the intracell and intercell hopping integrals, respectively.
Fourier transforming to crystal momentum space gives a two-band model
\begin{align}
H(k) = \begin{pmatrix}
0 & t_1 + t_2 e^{-ik} \\
t_1 + t_2 e^{ik} & 0
\end{pmatrix}
= \vec{d}(k) \cdot \vec{\sigma},
\end{align}
where \( \vec{d}(k) = (t_1 + t_2 \cos k, t_2 \sin k, 0) \) is the pseudo-spin vector which lies on the $x,y$ plane and \( \vec{\sigma} \) is the vector of Pauli matrices in sublattice space.
The energy bands are given by:
\begin{align}
\varepsilon_\pm(k) = \pm |\vec{d}(k)| = \pm \sqrt{t_1^2 + t_2^2 + 2t_1t_2 \cos k},
\end{align}
and we assume that the lower (valence band) is 
completely filled at $T = 0K$. 
A sketch of the model and resulting band-structure is given in Fig.~\ref{fig:SSHModel}(a).

The model exhibits two distinct phases.
In the trivial phase, specified by (\( |t_1| > |t_2| \)),  the bulk bands are gapped and the Hamiltonian trajectory in \( \vec{d}(k) \)-space does not encircle the origin.
However, in the topological phase (\( |t_2| > |t_1| \)) the bulk bands remain gapped, but the winding of \( \vec{d}(k) \) around the origin is nontrivial, signaling a topological phase, as shown in Fig.~\ref{fig:SSHModel}(b)
The topology of the SSH model is characterized by the Zak phase, defined as the Berry phase accumulated over the Brillouin zone for the occupied band,
\begin{align}
\gamma = i \int_{\text{BZ}} \langle u_k | \partial_k u_k \rangle\, dk,
\end{align}
where \( |u_k\rangle \) is the periodic part of the Bloch wavefunction. In the topological phase, \( \gamma = \pi \), while in the trivial phase \( \gamma = 0 \) (modulo \( 2\pi \)). This quantized geometric phase reflects the underlying winding number and gives rise to robust edge states under open boundary conditions.
The Zak phase influences selection rules, phase-sensitive transport, and the symmetry of nonlinear signals. Optical probes sensitive to coherence and phase evolution—such as high-harmonic generation, 2D spectroscopy, and ultrafast pump-probe experiments—can reveal signatures of the topological character through their spectral asymmetries and time-resolved interference patterns. The SSH model 
has inversion symmetry in both the trivial and topological 
phases and the Zak phase is quantized to 0  or $\pi$ precisely because of this symmetry~\cite{Zak1989}.

We add to our model the
 long-range coupling between electrons and longitudinal optical (LO) phonons in polar materials.
Within the Migdal approximation~\cite{Migdal1958,Roy2014}, the electron self-energy due to Fr{\"o}hlich coupling~\cite{Frohlich1950}
at temperature $T$ is given by
\begin{align}
\Sigma(\omega) = \int \frac{d^d q}{(2\pi)^d} |M(\mathbf{q})|^2 \left[ \frac{n_{\text{ph}} + 1}{\omega - \omega_{\text{LO}} - \epsilon_{\mathbf{k+q}} + i\delta} + \frac{n_{\text{ph}}}{\omega + \omega_{\text{LO}} - \epsilon_{\mathbf{k+q}} + i\delta} \right],
\end{align}
where $n_{\text{ph}} = \left( e^{\omega_{\text{LO}}/T} - 1 \right)^{-1}$ is the Bose-Einstein phonon occupation number.
In tight-binding models such as the SSH chain, one typically assumes that optical phonons are dispersionless, $\omega_{\mathbf{q}} = \omega_{\text{LO}}$,
and the coupling $M(\mathbf{q})$ is approximated by a constant or smooth function localized in $q$-space.
This gives us a compact form of the self-energy:
\begin{align}
\Sigma(\omega) \approx \lambda^2 \left[ \frac{n_{\text{ph}} + 1}{\omega - \omega_{\text{LO}} + i\delta} + \frac{n_{\text{ph}}}{\omega + \omega_{\text{LO}} + i\delta} \right],
\end{align}
where $\lambda$ is an effective coupling constant encapsulating the strength of the electron–phonon interaction.
This form can be included in Green’s function-based computations of nonlinear response functions (e.g., third-order susceptibilities or 2D spectra) 
such that the retarded 
Green's functions become
\begin{align}
\tilde G^{\mathrm{ret}}(\omega, k) = \frac{1}{ \omega + i\eta - H_{\text{SSH}}(k) - \Sigma(\omega) }
\end{align}
In our third-order calculations, we include the Fr\"ohlich self-energy $\Sigma(\omega)$ to model phonon-induced lifetime broadening and renormalization. For consistency, this correction should also enter the linear susceptibility $\chi^{(1)}$, but we omit it there to highlight the coherence-driven features of $\chi^{(3)}$. 
This is justifiable 
since our focus here is on topological signatures and phase coherence, rather than electron–phonon physics \textit{per se.}
Future work may incorporate a fully self-consistent treatment of $\Sigma(\omega)$ across all orders of response.

\begin{figure*}[ht]
    \centering
 \subfigure[]{\includegraphics[width=0.45\linewidth]{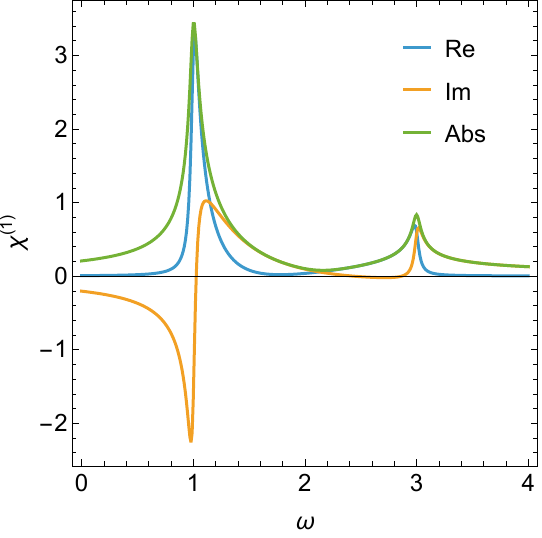}}
 \subfigure[]{   \includegraphics[width=0.45\linewidth]{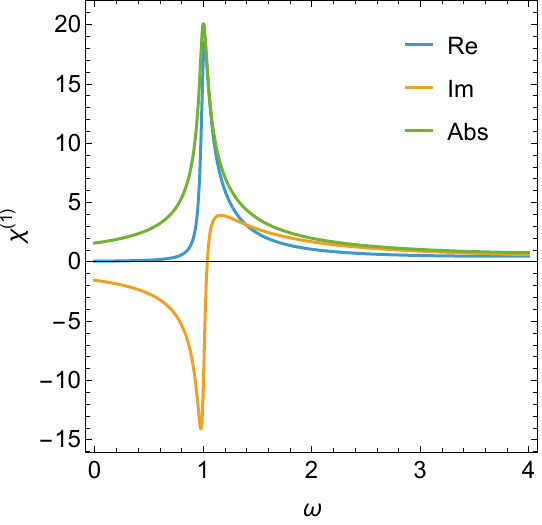}}
    \caption{Linear response of SSH model in the (a) normal regime ($t_2 = t_1/2$) and (b) 
    topological regime ($t_2 =1.5 t_1$).
    }
    \label{fig:2}
\end{figure*}

\subsubsection{Linear Response}
For the case of solid-state systems,
$\rho_{\text{eq}}$ contains the electronic populations given by the Fermi-Dirac distribution
and the trace becomes an integral over the 
BZ.
Accordingly, the current susceptibility tensor
in the velocity gauge can be written in the Bloch basis as
\begin{align}
\chi^{(1)}_{\mu \nu}(\omega)
&=\chi^{(1)}(\omega) = \text{Tr}[\hat{\jmath}_\mu 
\tilde G(\omega) \hat{\jmath}_\nu \tilde G(\omega)] \\
&= \frac{i}{\hbar} \sum_{k} \sum_{m \neq n} \frac{f_n(k) - f_m(k)}{\hbar \omega - \varepsilon_m(k) + \varepsilon_n(k) + i\eta}
\left\langle u_n(k) \left| \hat{\jmath}_\mu(k) \right| u_m(k) \right\rangle
\left\langle u_m(k) \left| \hat{\jmath}_\nu(k) \right| u_n(k) \right\rangle,
\end{align}
where \( f_n(k) \) is the Fermi–Dirac occupation factor, and \( \hat{\jmath}_\mu(k) = \frac{1}{\hbar} \partial_{k_\mu} H(k) \) is the current operator.
Similarly, in the length gauge, the polarization susceptibility is instead given by
\begin{equation}
\chi^{(1)}_{\mathrm{P}, \mu \nu}(\omega) = \frac{i q^2}{\hbar} \sum_k \sum_{m \neq n} \frac{f_n(k) - f_m(k)}{\hbar \omega - \varepsilon_m(k) + \varepsilon_n(k) + i\eta} 
\left\langle u_n(k) \left| \hat{r}_\mu \right| u_m(k) \right\rangle 
\left\langle u_m(k) \left| \hat{r}_\nu \right| u_n(k) \right\rangle,
\end{equation}
with the position operator expressed in the Bloch basis as \( \hat{r}_\mu = i \partial_{k_\mu} \).
These two forms are related by the identity
\begin{equation}
\left\langle u_n(k) \left| \partial_{k_\mu} H(k) \right| u_m(k) \right\rangle 
= \left( \varepsilon_n(k) - \varepsilon_m(k) \right) \left\langle u_n(k) \left| \partial_{k_\mu} \right| u_m(k) \right\rangle,
\quad \text{for } n \neq m,
\end{equation}
which ensures gauge equivalence when the full band structure is included. In numerical calculations, maintaining this consistency is critical for preserving gauge invariance.
We emphasize that while the velocity gauge emphasizes dynamical coupling via band velocities, the length gauge captures geometric information encoded in the Berry connection. Both representations yield equivalent predictions when implemented with care, including under the rotating-wave approximation. 

\begin{figure}[h!]
    \subfigure[]{\includegraphics[width=0.4\textwidth]{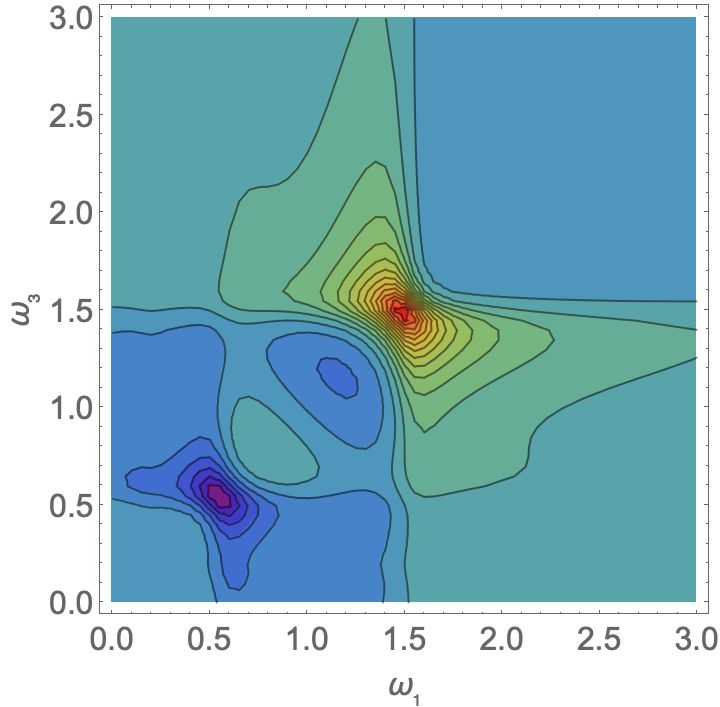}}
    \subfigure[]{\includegraphics[width=0.4\textwidth]{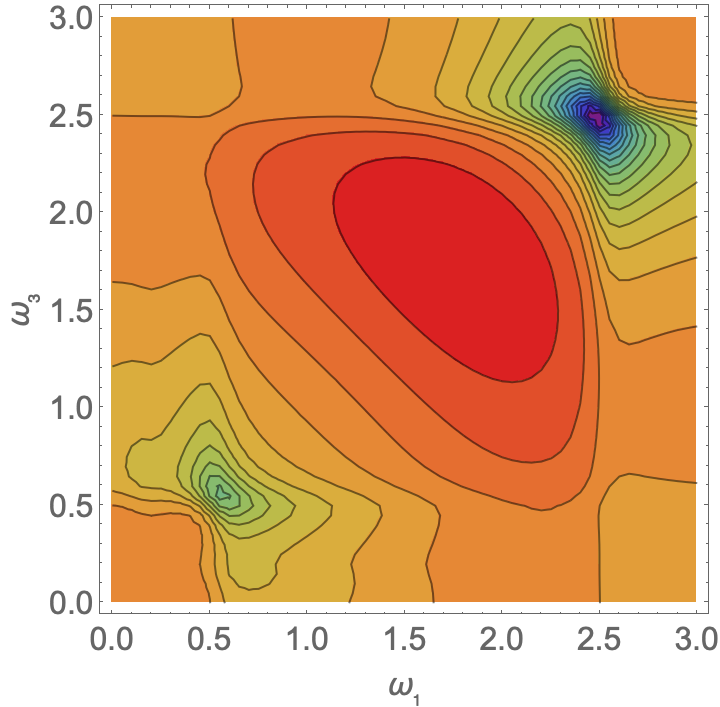}}
    \subfigure[]{\includegraphics[width=0.4\textwidth]{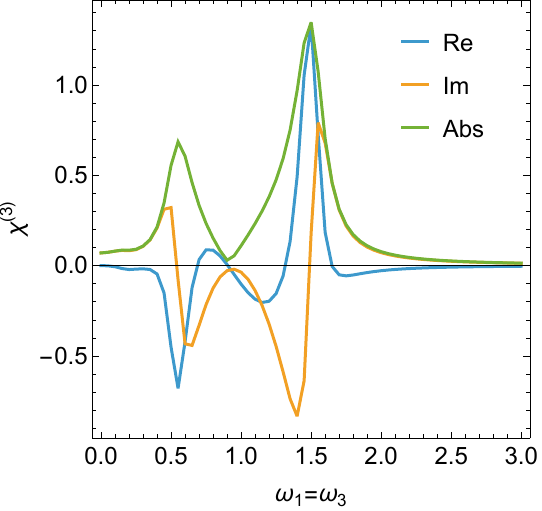}}
    \subfigure[]{\includegraphics[width=0.4\textwidth]{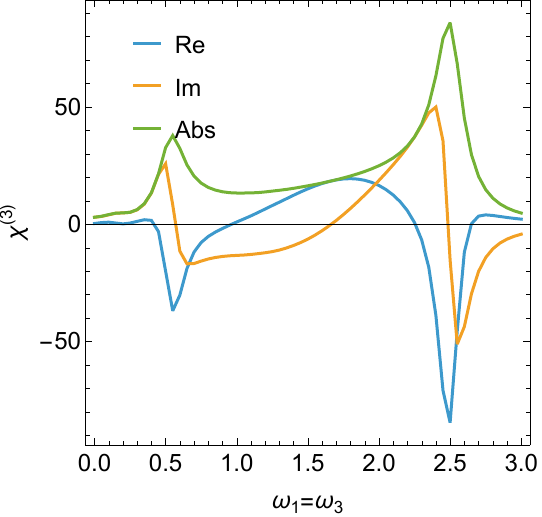}}
\caption{
Third-order nonlinear spectra for the SSH model in the trivial (left) and topological (right) phases. Panels (a,b) show the imaginary part (absorptive) of the rephasing $\chi^{(3)}(\omega_3,\omega_1)$ signal; panels (c,d) show the real part (dispersive). In the trivial phase ($t_2/t_1 < 1$), both low- and high-energy features appear with the same sign, while in the topological phase ($t_2/t_1 > 1$), the high-energy feature at $k = 0$ acquires an additional $\pi$ phase, reversing the sign of its real component. This inversion arises from the topological winding of the Bloch wavefunctions and indicates a geometric phase difference between the contributing interband coherence pathways. Spectral features align with half-gap energies at \( k = \pi \) and \( k = 0 \), reflecting resonance conditions where the sum of excitation and emission frequencies matches the interband transition energy, as expected for single-quantum coherence pathways in third-order response.
}
    \label{Fig:3}
\end{figure}

In the linear current response \( \chi^{(1)}(\omega) \) shown in Fig.~\ref{fig:2} we see a marked difference between the topologically trivial and nontrivial phases of the SSH model. In the trivial phase (\( |t_1| > |t_2| \)), the band structure exhibits prominent interband gaps at both \( k = \pi \) and \( k = 0 \), yielding two distinct absorption peaks corresponding to \( \omega = 2|t_1-t_2| \) and \( \omega = 2|t_1+t_2| \), respectively. These features arise from the joint density of states (JDOS) combined with sizable velocity matrix elements across both regions of the Brillouin zone. In contrast, the topological phase (\( |t_2| > |t_1| \)) supports only a single low-energy peak near \( k=\pi \), despite the presence of an optical gap at \( k = 0 \). This asymmetry is not due to changes in the energy spectrum alone, but rather to a suppression of the current operator matrix elements near \( k = 0 \), reflecting the altered wavefunction topology. The transition amplitude vanishes due to destructive interference in the Bloch wavefunction overlaps, a signature of the nontrivial Zak phase. Thus, the spectral selectivity of \( \chi^{(1)}(\omega) \) offers a direct probe of the underlying topological phase through both energy- and momentum-resolved optical response.

\subsubsection{Nonlinear Spectroscopy of the SSH Model}

To probe the nonlinear optical response of the SSH model, we compute the third-order current response function in both rephasing and non-rephasing configurations using the velocity-gauge formalism. Assuming the valence band is fully occupied and the conduction band is empty, the trace over the density matrix becomes an integral over the Brillouin zone and a sum over band indices. We define the rephasing and non-rephasing components as
\begin{align}
\chi^{(3)}_{\mathrm{R}}(\omega_3, \tau_2, \omega_1) &= \left(\frac{i}{\hbar}\right)^3 \int_{\mathrm{BZ}} \frac{dk}{2\pi} \mathrm{Tr} \left[
j(k) \tilde G(\omega_3, k) j(k)  G_2(\tau_2, k) j^\dagger(k) \tilde G(\omega_1, k) j^\dagger(k)
\right],
\nonumber\\
\chi^{(3)}_{\mathrm{NR}}(\omega_3, \tau_2, \omega_1) &= \left(\frac{i}{\hbar}\right)^3 \int_{\mathrm{BZ}} \frac{dk}{2\pi}  \mathrm{Tr} \left[
j^\dagger(k) \tilde G(\omega_3, k) j(k)  G_2(\tau_2, k)  j(k)  \tilde G(\omega_1, k) j^\dagger(k)
\right],
\end{align}
where the current operator is given by \( j(k) = \partial_k H(k) \), and \( G_2(\tau_2, k) \) describes time evolution during the population interval \( \tau_2 \). Since \( j^\dagger(k) = j(-k) \), these expressions preserve momentum conservation and correctly account for excitations and de-excitations on opposite sides of the energy bands.

The expressions for $\chi^{(3)}_{\mathrm{NR/R}}(\omega_3, \tau_2, \omega_1)$ involve products of the current operator, retarded Green’s functions, and evolution operators and correspond to specific Liouville pathways with fixed operator ordering. These may be viewed as contractions of the fully causal, commutator-structured response
\[
\chi^{(3)}(\omega_3, \tau_2, \omega_1) = \left( \frac{i}{\hbar} \right)^3 \langle [[[j(\omega_3), j_H(\tau_2)], j(\omega_1)], j(0)] \rangle,
\]
where the population delay \( \tau_2 \) is retained in real time.  The non-commutator expressions used above correspond to particular time orderings and selection of a single pathway (e.g., ESA or rephasing) from the full set encoded by the commutator form.
This is a gauge-consistent, pathway-resolved projection of the full nonlinear susceptibility.

Fig.~\ref{Fig:3} shows the third-order rephasing responses
for both the trivial ($t_1=1,t_2=0.5$) and topological ($t_1 =1,t_2=1.5$) phases of the model. The bottom two panels
show diagonal slices according to $\omega_1 = \omega_3$ 
for a fixed population time $\tau_2 =3$ in natural units. 
 In a clean, translationally invariant SSH model, we find nearly indistinguishable spectra for rephasing and non-rephasing signals due to the absence of inhomogeneous broadening.
The representation works for the SSH model since the
conduction and valence bands are symmetric; however, the 
transitions will be shifted to $1/2$ the bandwidth at a given $k$.  

\begin{figure}[h]
    \centering
    \includegraphics[width=0.5\linewidth]{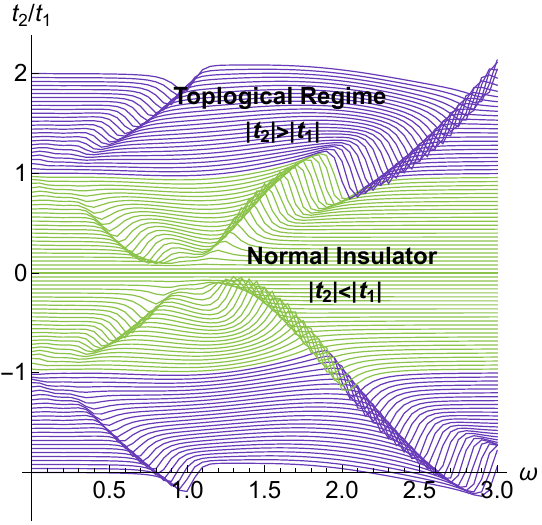}
\caption{
Waterfall plot of diagonal slices through the third-order rephasing spectrum $\chi^{(3)}(\omega_1=\omega_3)$ as a function of the intercell hopping $t_2$ at fixed $t_1 = 1$. Each trace corresponds to a constant-$t_2$ spectrum offset vertically for clarity. As $t_2$ increases across the topological transition ($t_2/t_1 = 1$), the relative phase of the high-energy feature at $k = 0$ changes sign. This smooth but distinct inversion is a spectroscopic signature of the topological transition and reflects the momentum-dependent phase structure of the current matrix elements. Broadening from Fr\"ohlich self-energy and lifetime effects smooths the crossover, but this phase shift 
remains a robust spectroscopic marker of the topological transition.
}
    \label{Fig:JoyDivision}
\end{figure}

Figure~\ref{Fig:JoyDivision} illustrates this transition via diagonal slices through the third-order rephasing spectra as a function of intercell hopping \( t_2 \). In the trivial phase, both low- and high-energy peaks show absorptive lineshapes. As the system enters the topological regime, the high-energy contribution—originating from transitions near \( k = 0 \)—acquires a relative phase of \( \pi \), reversing the sign of its real part. This spectral inversion reflects the change in winding of \( \hat{d}(k) \), as \( j(k) \propto \hat{d}(k) \times \partial_k \hat{d}(k) \) becomes out of phase at \( k = 0 \) and \( k = \pi \).

Finally, the relative intensity of the spectral peaks depends on the joint density of states (JDOS), current matrix elements, and phonon-induced damping. Near \( k = 0 \), the flatter band structure enhances JDOS and increases \( j(k) \), favoring strong transitions. Near \( k = \pi \), steeper dispersion reduces state density and suppresses the interband coupling, especially in the presence of symmetry constraints. Furthermore, temperature-dependent broadening from the Fr\"ohlich self-energy asymmetrically attenuates low-energy transitions. Together, these effects explain the dominance of the high-energy peak and the interference-driven redistribution of spectral weight across the topological transition.

\section{Discussion}

We have presented a velocity-gauge formalism for computing linear and nonlinear current responses in topological tight-binding models, using the SSH chain as a representative example. By retaining the full minimal coupling Hamiltonian and avoiding the rotating wave approximation, we ensured gauge consistency and captured the full temporal structure of the response. Linear spectra reflect the band structure and density of states, while third-order signals reveal more subtle features arising from interband coherence and phase winding. In particular, we observed a phase-sensitive inversion of spectral features across the topological transition, illustrating how nonlinear optical response functions can serve as probes of geometric and topological character in extended systems~\cite{Mukamel1995,Kruk2022}.

A key aspect of our approach is the omission of the rotating wave approximation (RWA), which is often employed to simplify nonlinear optical calculations by retaining only near-resonant terms. Instead, we compute the full current-based response by evaluating time-ordered correlation functions using retarded Green’s functions, thereby retaining both resonant and counter-rotating contributions. This formulation preserves the complete temporal structure of the signal and ensures gauge consistency. The inclusion of all quantum pathways is especially important in systems with nontrivial band topology, where virtual interband coherences and off-resonant transitions contribute significantly to the nonlinear response. Our results show that such contributions are essential to capturing phase-sensitive features and spectral asymmetries tied to geometric and topological structure.

Although linear response functions are gauge invariant under exact treatment, nonlinear susceptibilities generally are not—particularly when calculated in truncated bases. In the velocity gauge, the field couples via the vector potential, leading to expressions involving products of current operators and Green’s functions~\cite{AversaSipe1995,SipeShkrebtii2000}. The length gauge, by contrast, couples directly to the electric field through the position operator and introduces Berry connection terms. In principle, both approaches yield identical results in a complete Hilbert space; however, in practical implementations with a finite number of bands, this equivalence breaks down~\cite{Wismer2018,Ventra2004}. As a result, third-order spectra can differ significantly between gauges, especially if interband and intraband contributions are not treated on equal footing. In our analysis of the two-band SSH model, we adopt a current-based formalism consistent with the velocity gauge. While this framework captures key topological features encoded in interband coherence pathways, any quantitative comparison with length-gauge calculations would require careful inclusion of gauge-restoring terms. Gauge-dependent discrepancies in lineshapes or intensities should therefore be interpreted with caution in reduced models.

In the linear response, the absorption peak tracks the direct bandgap at $ka = \pi$, where both the joint density of states and current matrix elements are maximal. In contrast, the third-order non-rephasing spectrum exhibits peaks at approximately half the bandgap energies associated with $ka = \pi$ and $ka = 0$. This reflects resonance conditions characteristic of single-quantum coherence pathways, where each field interaction contributes fractionally to the total transition energy. These features emerge prominently in the topological phase and reflect coherent mixing between interband channels enabled by the nontrivial winding of the Bloch wavefunctions. The third-order response thus encodes both energy differences and phase coherence, offering sensitivity to topological structure that is absent in linear spectroscopy~\cite{Parker2019,Fregoso2018}.

In our third-order response calculations, we evaluate expressions of the form
\begin{equation}
\chi^{(3)}(\omega_1,\tau_2, \omega_3) \sim \text{Tr}\left[ j G^R(\omega_3) j G_2(\tau_2) j^\dagger G^R(\omega_1) j^\dagger \right].
\end{equation}
This represents a prototypical rephasing contribution to \( \chi^{(3)} \), formulated in Hilbert space using time-ordered current insertions and retarded Green’s functions. Although this form does not explicitly resolve all Liouville pathways, it effectively captures the coherent evolution of an interband excitation through its response to multiple frequency components. The resonance structure observed in diagonal frequency slices thus reflects energy-sharing between field interactions, rather than direct transitions at a single Green's function pole.

Because our formalism constructs the third-order response entirely within Hilbert space using retarded Green's functions, the density matrix evolution is implicitly encoded in the sequence of operator insertions and propagators, without explicit tracking of ket- and bra-side evolution or resolution of individual Liouville pathways such as ground-state bleach (GSB), stimulated emission (SE), or excited-state absorption (ESA). In Liouville-space response theory, these contributions are defined by time-contour-ordered interaction sequences with distinct coherence and population intervals, often resolved via double-sided Feynman diagrams. In contrast, the Hilbert-space approach captures these effects indirectly through the symmetric frequency dependence of time-ordered current insertions. This structure naturally gives rise to resonances at \( \omega \approx \Delta_k/2 \) in diagonal frequency slices \( \omega_1 = \omega_3 = \omega \), as energy is effectively shared across the initial and final interactions. While no true poles exist at half the bandgap, the Hilbert-space sequence produces apparent resonances due to distributed coherence evolution in the complex frequency plane.

To interpret this physically, it is helpful to recall that the third-order signal in the rephasing configuration reflects the coherent evolution of an interband excitation — a valence band hole and a conduction band electron at the same momentum $k$. In this picture, the system is initially excited into a coherence between the ground state and an interband excited state $|c_k\rangle \langle v_k|$, which evolves under the Hamiltonian as $\rho_{cv}(t) \sim e^{-i (E_c(k) - E_v(k)) t}$. However, in the frequency-domain formulation of $\chi^{(3)}(\omega_1, \omega_3)$, this excitation is distributed across two frequency arguments. For diagonal slices with $\omega_1 = \omega_3 = \omega$, the resonance condition becomes $2\omega \approx \Delta_k$, corresponding to the case where each field interaction shares half the transition energy. This does not imply a true pole at $\Delta_k/2$, but arises from how energy is partitioned across time-ordered interactions in the nonlinear pathway.
These pathways naturally appear in rephasing and non-rephasing configurations alike and explain the consistent appearance of peaks near half the gap across both topological and trivial regimes.
Thus, the half-gap spectral features are a direct consequence of the structure of $\chi^{(3)}(\omega_1, \omega_3)$ and should not be interpreted as representing shifted single-particle excitations. Instead, they encode the internal coherence structure of the system and reflect the resonance conditions of coherent multiphoton pathways.

\section{Perspective}

Our results establish a velocity-gauge framework for nonlinear current response that is both gauge-consistent and sensitive to the topological structure of Bloch bands. By applying this approach to the SSH model, we demonstrated how third-order optical spectra encode not only band energies and joint density of states but also interband phase coherence and geometric winding. The appearance of phase reversals and interference effects in the nonlinear response provides a clear spectroscopic fingerprint of topological character, even in minimal two-band models. These findings support a broader view that nonlinear optical observables—especially those involving time-ordered pathways and off-resonant couplings—can serve as precise, experimentally accessible probes of quantum geometry in solids and synthetic matter. Looking ahead, our formalism may be extended to disordered systems, finite temperatures, and driven phases, offering a versatile platform for exploring the interplay between symmetry, coherence, and topology in nonlinear light–matter interactions~\cite{Schlawin2018,Meier2016,Lu2014}.

The velocity-gauge framework we develop is not limited to the SSH model. For example, in the Rice--Mele model~\cite{RiceMele1982}, where a staggered on-site potential breaks inversion symmetry, the formalism remains valid and can be used to compute both bulk and surface nonlinear responses, including second-order susceptibilities that are symmetry-forbidden in the SSH case. 
Extensions of SSH-type models to real materials such as 1H-transition metal dichalcogenides reveal higher-order topological phases with protected edge and corner modes, enabling spectroscopic access to topological order beyond simple band inversion~\cite{Zeng2021}. 

Similarly, for two-dimensional systems like the Haldane model~\cite{Haldane1988}---featuring broken time-reversal symmetry and a nonzero Chern number---the same Green's function--based approach applies, with the current operator defined in terms of momentum derivatives of the multi-band Bloch Hamiltonian. These extensions allow exploration of Berry curvature, Chern numbers, and topological edge states via nonlinear optical observables. Because the formalism is based on current--current correlation functions and time-ordered perturbation theory, it naturally scales to systems with more bands, broken symmetries, and higher dimensions, making it a broadly applicable tool for probing quantum geometry in diverse lattice models. Extensions of this framework to anyonic systems suggest that velocity-gauge response theory may also provide a natural platform for probing fractional statistics and non-Abelian coherence effects in driven, low-dimensional systems~\cite{bittner2025statistical}.
Finally, experimental studies have uncovered hidden breathing kagome topology in hexagonal transition metal dichalcogenides, opening new avenues for nonlinear probes of fragile and higher-order topological phases~\cite{Jung2022}.

We have recently reported complex coherent spectral lineshapes that reveal strong multiparticle correlations in optical microcavities in the strong coupling limit~\cite{quiroscordero2025resolving}. We identify spectral structure of polariton-polariton and polariton-exciton scattering, and we consider that the formalism discussed here will reveal the many-body quantum dynamics represented by the rich multidimentional coherent lineshapes in this system. 

We have also shown that correlated dissipation can induce spontaneous quantum synchronization in coupled oscillators~\cite{Tyagi2024JPCL,Bittner2025JCP}, revealing that noise can enhance coherence and lead to emergent phase locking. These results parallel the current framework, where coherence pathways in the third-order response encode geometric and topological structure. A natural extension of the present work would be to examine how dissipation-induced synchronization manifests in topological lattices and whether spectroscopic signatures such as those observed here may track dynamical phase transitions in open systems.

\section{Biographies}

\textbf{Eric R.\ Bittner} is the Hugh Roy and Lillie Cranz Cullen Distinguished Professor of Physics at the University of Houston. He received his Ph.D. in Chemistry from the University of Chicago in 1994. His research focuses on quantum dynamics, nonlinear spectroscopy, and topological phenomena in molecular and condensed-phase systems.

\noindent \textbf{Carlos Silva-Acu\~{n}a} holds the Canada Excellence Research Chair in Light-Matter Interactions, is Professor of Physics, Director of the Institut Courtois, and the Institut Courtois Director's Research Chair at the Universit\'e de Montr\'eal. He received a PhD in Chemical Physics from the University of Minnesota in 1998. His research focuses on ultrafast coherent spectroscopy of materials, focusing on organic and hybrid semiconductors, photonic structures, and strongly correlated quantum materials.
\begin{acknowledgement}

CSA acknowledges funding from the Government of Canada (Canada Excellence Research Chair CERC-2022-00055) and support from the Institut Courtois, Facult\'e des arts et des sciences, Universit\'e de Montr\'eal (Chaire de Recherche de l'Institut Courtois).
ERB was supported by the National Science Foundation (CHE-2404788) and the Robert A. Welch Foundation (E-1337).

\end{acknowledgement}

\section*{Data Availability}

All data supporting the findings of this study --- including numerical simulations, spectra, and analysis code --- are publicly available on the Borealis Dataverse Repository [Link to be inserted prior to publication]. A Mathematica notebook is included to reproduce the nonlinear response calculations and band-structure analysis.
This notebook is freely available on the Wolfram Cloud via 
\href{https://tinyurl.com/2myfznc9}{https://tinyurl.com/2myfznc9}.


\bibliography{references}

\providecommand{\latin}[1]{#1}
\makeatletter
\providecommand{\doi}
  {\begingroup\let\do\@makeother\dospecials
  \catcode`\{=1 \catcode`\}=2 \doi@aux}
\providecommand{\doi@aux}[1]{\endgroup\texttt{#1}}
\makeatother
\providecommand*\mcitethebibliography{\thebibliography}
\csname @ifundefined\endcsname{endmcitethebibliography}
  {\let\endmcitethebibliography\endthebibliography}{}
\begin{mcitethebibliography}{48}
\providecommand*\natexlab[1]{#1}
\providecommand*\mciteSetBstSublistMode[1]{}
\providecommand*\mciteSetBstMaxWidthForm[2]{}
\providecommand*\mciteBstWouldAddEndPuncttrue
  {\def\EndOfBibitem{\unskip.}}
\providecommand*\mciteBstWouldAddEndPunctfalse
  {\let\EndOfBibitem\relax}
\providecommand*\mciteSetBstMidEndSepPunct[3]{}
\providecommand*\mciteSetBstSublistLabelBeginEnd[3]{}
\providecommand*\EndOfBibitem{}
\mciteSetBstSublistMode{f}
\mciteSetBstMaxWidthForm{subitem}{(\alph{mcitesubitemcount})}
\mciteSetBstSublistLabelBeginEnd
  {\mcitemaxwidthsubitemform\space}
  {\relax}
  {\relax}

\bibitem[Mukamel(1995)]{Mukamel1995}
Mukamel,~S. \emph{Principles of Nonlinear Optical Spectroscopy}; Oxford
  University Press, 1995\relax
\mciteBstWouldAddEndPuncttrue
\mciteSetBstMidEndSepPunct{\mcitedefaultmidpunct}
{\mcitedefaultendpunct}{\mcitedefaultseppunct}\relax
\EndOfBibitem
\bibitem[Li \latin{et~al.}(2023)Li, Lomsadze, Moody, Smallwood, and
  Cundiff]{li2023optical}
Li,~H.; Lomsadze,~B.; Moody,~G.; Smallwood,~C.; Cundiff,~S. \emph{Optical
  multidimensional coherent spectroscopy}; Oxford University Press, 2023\relax
\mciteBstWouldAddEndPuncttrue
\mciteSetBstMidEndSepPunct{\mcitedefaultmidpunct}
{\mcitedefaultendpunct}{\mcitedefaultseppunct}\relax
\EndOfBibitem
\bibitem[Fresch \latin{et~al.}(2023)Fresch, Camargo, Shen, Bellora, Pullerits,
  Engel, Cerullo, and Collini]{fresch2023two}
Fresch,~E.; Camargo,~F.~V.; Shen,~Q.; Bellora,~C.~C.; Pullerits,~T.;
  Engel,~G.~S.; Cerullo,~G.; Collini,~E. Two-dimensional electronic
  spectroscopy. \emph{Nature Reviews Methods Primers} \textbf{2023}, \emph{3},
  84\relax
\mciteBstWouldAddEndPuncttrue
\mciteSetBstMidEndSepPunct{\mcitedefaultmidpunct}
{\mcitedefaultendpunct}{\mcitedefaultseppunct}\relax
\EndOfBibitem
\bibitem[Kelley(2022)]{kelley2022condensed}
Kelley,~A.~M. \emph{Condensed-phase molecular spectroscopy and photophysics};
  John Wiley \& Sons, 2022\relax
\mciteBstWouldAddEndPuncttrue
\mciteSetBstMidEndSepPunct{\mcitedefaultmidpunct}
{\mcitedefaultendpunct}{\mcitedefaultseppunct}\relax
\EndOfBibitem
\bibitem[Klingshirn(2012)]{klingshirn2012semiconductor}
Klingshirn,~C.~F. \emph{Semiconductor optics}; Springer Science \& Business
  Media, 2012\relax
\mciteBstWouldAddEndPuncttrue
\mciteSetBstMidEndSepPunct{\mcitedefaultmidpunct}
{\mcitedefaultendpunct}{\mcitedefaultseppunct}\relax
\EndOfBibitem
\bibitem[Hamm and Zanni(2011)Hamm, and Zanni]{hamm2011concepts}
Hamm,~P.; Zanni,~M. \emph{Concepts and methods of 2D infrared spectroscopy};
  Cambridge University Press, 2011\relax
\mciteBstWouldAddEndPuncttrue
\mciteSetBstMidEndSepPunct{\mcitedefaultmidpunct}
{\mcitedefaultendpunct}{\mcitedefaultseppunct}\relax
\EndOfBibitem
\bibitem[Frisk~Kockum \latin{et~al.}(2019)Frisk~Kockum, Miranowicz,
  De~Liberato, Savasta, and Nori]{frisk2019ultrastrong}
Frisk~Kockum,~A.; Miranowicz,~A.; De~Liberato,~S.; Savasta,~S.; Nori,~F.
  Ultrastrong coupling between light and matter. \emph{Nature Reviews Physics}
  \textbf{2019}, \emph{1}, 19--40\relax
\mciteBstWouldAddEndPuncttrue
\mciteSetBstMidEndSepPunct{\mcitedefaultmidpunct}
{\mcitedefaultendpunct}{\mcitedefaultseppunct}\relax
\EndOfBibitem
\bibitem[Forn-D{\'\i}az \latin{et~al.}(2019)Forn-D{\'\i}az, Lamata, Rico, Kono,
  and Solano]{forn2019ultrastrong}
Forn-D{\'\i}az,~P.; Lamata,~L.; Rico,~E.; Kono,~J.; Solano,~E. Ultrastrong
  coupling regimes of light-matter interaction. \emph{Reviews of Modern
  Physics} \textbf{2019}, \emph{91}, 025005\relax
\mciteBstWouldAddEndPuncttrue
\mciteSetBstMidEndSepPunct{\mcitedefaultmidpunct}
{\mcitedefaultendpunct}{\mcitedefaultseppunct}\relax
\EndOfBibitem
\bibitem[Le~Boit{\'e}(2020)]{le2020theoretical}
Le~Boit{\'e},~A. Theoretical methods for ultrastrong light--matter
  interactions. \emph{Advanced Quantum Technologies} \textbf{2020}, \emph{3},
  1900140\relax
\mciteBstWouldAddEndPuncttrue
\mciteSetBstMidEndSepPunct{\mcitedefaultmidpunct}
{\mcitedefaultendpunct}{\mcitedefaultseppunct}\relax
\EndOfBibitem
\bibitem[Stokes and Nazir(2021)Stokes, and Nazir]{stokes2021ultrastrong}
Stokes,~A.; Nazir,~A. Ultrastrong time-dependent light-matter interactions are
  gauge relative. \emph{Physical Review Research} \textbf{2021}, \emph{3},
  013116\relax
\mciteBstWouldAddEndPuncttrue
\mciteSetBstMidEndSepPunct{\mcitedefaultmidpunct}
{\mcitedefaultendpunct}{\mcitedefaultseppunct}\relax
\EndOfBibitem
\bibitem[Cohen-Tannoudji \latin{et~al.}(1992)Cohen-Tannoudji, Dupont-Roc, and
  Grynberg]{Cohen_Tannoudji_atomphoton}
Cohen-Tannoudji,~C.; Dupont-Roc,~J.; Grynberg,~G. \emph{Atom-Photon
  Interactions: Basic Processes and Applications}; Wiley: New York, 1992\relax
\mciteBstWouldAddEndPuncttrue
\mciteSetBstMidEndSepPunct{\mcitedefaultmidpunct}
{\mcitedefaultendpunct}{\mcitedefaultseppunct}\relax
\EndOfBibitem
\bibitem[Tanaka and Takayoshi(2015)Tanaka, and Takayoshi]{tanaka2015short}
Tanaka,~A.; Takayoshi,~S. A short guide to topological terms in the effective
  theories of condensed matter. \emph{Science and Technology of Advanced
  Materials} \textbf{2015}, \emph{16}, 014404\relax
\mciteBstWouldAddEndPuncttrue
\mciteSetBstMidEndSepPunct{\mcitedefaultmidpunct}
{\mcitedefaultendpunct}{\mcitedefaultseppunct}\relax
\EndOfBibitem
\bibitem[Wang and Zhang(2017)Wang, and Zhang]{wang2017topological}
Wang,~J.; Zhang,~S.-C. Topological states of condensed matter. \emph{Nature
  materials} \textbf{2017}, \emph{16}, 1062--1067\relax
\mciteBstWouldAddEndPuncttrue
\mciteSetBstMidEndSepPunct{\mcitedefaultmidpunct}
{\mcitedefaultendpunct}{\mcitedefaultseppunct}\relax
\EndOfBibitem
\bibitem[Popelier(2016)]{popelier2016quantum}
Popelier,~P.~L. Quantum chemical topology. \emph{The chemical bond II: 100
  years old and getting stronger} \textbf{2016}, 71--117\relax
\mciteBstWouldAddEndPuncttrue
\mciteSetBstMidEndSepPunct{\mcitedefaultmidpunct}
{\mcitedefaultendpunct}{\mcitedefaultseppunct}\relax
\EndOfBibitem
\bibitem[Kumar \latin{et~al.}(2020)Kumar, Guin, Manna, Shekhar, and
  Felser]{kumar2020topological}
Kumar,~N.; Guin,~S.~N.; Manna,~K.; Shekhar,~C.; Felser,~C. Topological quantum
  materials from the viewpoint of chemistry. \emph{Chemical Reviews}
  \textbf{2020}, \emph{121}, 2780--2815\relax
\mciteBstWouldAddEndPuncttrue
\mciteSetBstMidEndSepPunct{\mcitedefaultmidpunct}
{\mcitedefaultendpunct}{\mcitedefaultseppunct}\relax
\EndOfBibitem
\bibitem[Cano and Bradlyn(2021)Cano, and Bradlyn]{cano2021band}
Cano,~J.; Bradlyn,~B. Band representations and topological quantum chemistry.
  \emph{Annual Review of Condensed Matter Physics} \textbf{2021}, \emph{12},
  225--246\relax
\mciteBstWouldAddEndPuncttrue
\mciteSetBstMidEndSepPunct{\mcitedefaultmidpunct}
{\mcitedefaultendpunct}{\mcitedefaultseppunct}\relax
\EndOfBibitem
\bibitem[Bradlyn \latin{et~al.}(2017)Bradlyn, Elcoro, Cano, Vergniory, Wang,
  Felser, Aroyo, and Bernevig]{Bradlyn2017}
Bradlyn,~B.; Elcoro,~L.; Cano,~J.; Vergniory,~M.~G.; Wang,~Z.; Felser,~C.;
  Aroyo,~M.~I.; Bernevig,~B.~A. Topological Quantum Chemistry. \emph{Nature}
  \textbf{2017}, \emph{547}, 298--305\relax
\mciteBstWouldAddEndPuncttrue
\mciteSetBstMidEndSepPunct{\mcitedefaultmidpunct}
{\mcitedefaultendpunct}{\mcitedefaultseppunct}\relax
\EndOfBibitem
\bibitem[Reiss(2000)]{Reiss2000StrongFieldGauge}
Reiss,~H.~R. Theoretical Methods in Strong-Field Laser Physics. \emph{Phys.
  Rev. A} \textbf{2000}, \emph{63}, 013409\relax
\mciteBstWouldAddEndPuncttrue
\mciteSetBstMidEndSepPunct{\mcitedefaultmidpunct}
{\mcitedefaultendpunct}{\mcitedefaultseppunct}\relax
\EndOfBibitem
\bibitem[Chelkowski and Bandrauk(2005)Chelkowski, and
  Bandrauk]{Chelkowski2005HHGGauge}
Chelkowski,~S.; Bandrauk,~A.~D. Gauge Effects in High-Order Harmonic
  Generation. \emph{Phys. Rev. A} \textbf{2005}, \emph{71}, 053815\relax
\mciteBstWouldAddEndPuncttrue
\mciteSetBstMidEndSepPunct{\mcitedefaultmidpunct}
{\mcitedefaultendpunct}{\mcitedefaultseppunct}\relax
\EndOfBibitem
\bibitem[Krausz and Ivanov(2009)Krausz, and
  Ivanov]{Krausz2009AttosecondScience}
Krausz,~F.; Ivanov,~M. Attosecond Physics. \emph{Rev. Mod. Phys.}
  \textbf{2009}, \emph{81}, 163--234\relax
\mciteBstWouldAddEndPuncttrue
\mciteSetBstMidEndSepPunct{\mcitedefaultmidpunct}
{\mcitedefaultendpunct}{\mcitedefaultseppunct}\relax
\EndOfBibitem
\bibitem[Hofmann \latin{et~al.}(2019)Hofmann, Hanus, Haessler,
  Trallero-Herrero, Murnane, Kapteyn, and Cirelli]{Hofmann2019NonDipoleHHG}
Hofmann,~C.; Hanus,~V.; Haessler,~S.; Trallero-Herrero,~C.~A.; Murnane,~M.~M.;
  Kapteyn,~H.~C.; Cirelli,~C. Non-Dipole Contributions to High-Order Harmonic
  Generation in the Long-Wavelength Limit. \emph{Nat. Commun.} \textbf{2019},
  \emph{10}, 703\relax
\mciteBstWouldAddEndPuncttrue
\mciteSetBstMidEndSepPunct{\mcitedefaultmidpunct}
{\mcitedefaultendpunct}{\mcitedefaultseppunct}\relax
\EndOfBibitem
\bibitem[Liu \latin{et~al.}(2020)Liu, Taguchi, B{\"u}rkle, Sato, Rubio, and
  De~Giovannini]{Liu2020trARPESGauge}
Liu,~H.; Taguchi,~K.; B{\"u}rkle,~M.; Sato,~S.~A.; Rubio,~A.; De~Giovannini,~U.
  Velocity-Gauge Real-Time TDDFT for Strong-Field Photoemission. \emph{Phys.
  Rev. B} \textbf{2020}, \emph{101}, 035110\relax
\mciteBstWouldAddEndPuncttrue
\mciteSetBstMidEndSepPunct{\mcitedefaultmidpunct}
{\mcitedefaultendpunct}{\mcitedefaultseppunct}\relax
\EndOfBibitem
\bibitem[De~Giovannini \latin{et~al.}(2016)De~Giovannini, H{\"u}bener, and
  Rubio]{DeGiovannini2016VelocityGaugeARPES}
De~Giovannini,~U.; H{\"u}bener,~H.; Rubio,~A. A Gauge-Invariant Approach to
  Quantum Electrodynamics in Solids. \emph{Nat. Commun.} \textbf{2016},
  \emph{7}, 12463\relax
\mciteBstWouldAddEndPuncttrue
\mciteSetBstMidEndSepPunct{\mcitedefaultmidpunct}
{\mcitedefaultendpunct}{\mcitedefaultseppunct}\relax
\EndOfBibitem
\bibitem[Aversa and Sipe(1995)Aversa, and Sipe]{AversaSipe1995}
Aversa,~C.; Sipe,~J.~E. Nonlinear optical susceptibilities of semiconductors:
  Results with a length-gauge analysis. \emph{Phys. Rev. B} \textbf{1995},
  \emph{52}, 14636--14645\relax
\mciteBstWouldAddEndPuncttrue
\mciteSetBstMidEndSepPunct{\mcitedefaultmidpunct}
{\mcitedefaultendpunct}{\mcitedefaultseppunct}\relax
\EndOfBibitem
\bibitem[Su \latin{et~al.}(1979)Su, Schrieffer, and Heeger]{Su1979}
Su,~W.~P.; Schrieffer,~J.~R.; Heeger,~A.~J. Solitons in Polyacetylene.
  \emph{Physical Review Letters} \textbf{1979}, \emph{42}, 1698--1701\relax
\mciteBstWouldAddEndPuncttrue
\mciteSetBstMidEndSepPunct{\mcitedefaultmidpunct}
{\mcitedefaultendpunct}{\mcitedefaultseppunct}\relax
\EndOfBibitem
\bibitem[Zak(1989)]{Zak1989}
Zak,~J. Berry's Phase for Energy Bands in Solids. \emph{Physical Review
  Letters} \textbf{1989}, \emph{62}, 2747--2750\relax
\mciteBstWouldAddEndPuncttrue
\mciteSetBstMidEndSepPunct{\mcitedefaultmidpunct}
{\mcitedefaultendpunct}{\mcitedefaultseppunct}\relax
\EndOfBibitem
\bibitem[Lu \latin{et~al.}(2014)Lu, Joannopoulos, and Soljačić]{Lu2014}
Lu,~L.; Joannopoulos,~J.~D.; Soljačić,~M. Topological photonics. \emph{Nature
  Photonics} \textbf{2014}, \emph{8}, 821--829\relax
\mciteBstWouldAddEndPuncttrue
\mciteSetBstMidEndSepPunct{\mcitedefaultmidpunct}
{\mcitedefaultendpunct}{\mcitedefaultseppunct}\relax
\EndOfBibitem
\bibitem[Meier \latin{et~al.}(2016)Meier, An, and Gadway]{Meier2016}
Meier,~E.~J.; An,~F.~A.; Gadway,~B. Observation of the topological soliton
  state in the Su–Schrieffer–Heeger model. \emph{Nature Communications}
  \textbf{2016}, \emph{7}, 13986\relax
\mciteBstWouldAddEndPuncttrue
\mciteSetBstMidEndSepPunct{\mcitedefaultmidpunct}
{\mcitedefaultendpunct}{\mcitedefaultseppunct}\relax
\EndOfBibitem
\bibitem[Barcikowski and Rudner(2021)Barcikowski, and Rudner]{Barcikowski2021}
Barcikowski,~S.; Rudner,~M.~S. Topological edge states in disordered
  Su–Schrieffer–Heeger chains. \emph{Physical Review B} \textbf{2021},
  \emph{103}, 195409\relax
\mciteBstWouldAddEndPuncttrue
\mciteSetBstMidEndSepPunct{\mcitedefaultmidpunct}
{\mcitedefaultendpunct}{\mcitedefaultseppunct}\relax
\EndOfBibitem
\bibitem[Migdal(1958)]{Migdal1958}
Migdal,~A.~B. Interaction between electrons and lattice vibrations in a normal
  metal. \emph{Soviet Physics JETP} \textbf{1958}, \emph{7}, 996--1001,
  Originally published in Zh. Eksp. Teor. Fiz. \textbf{34}, 1438--1446
  (1958)\relax
\mciteBstWouldAddEndPuncttrue
\mciteSetBstMidEndSepPunct{\mcitedefaultmidpunct}
{\mcitedefaultendpunct}{\mcitedefaultseppunct}\relax
\EndOfBibitem
\bibitem[Roy \latin{et~al.}(2014)Roy, Sau, and Sarma]{Roy2014}
Roy,~B.; Sau,~J.~D.; Sarma,~S.~D. Migdal's theorem and electron-phonon vertex
  corrections in Dirac materials. \emph{Physical Review B} \textbf{2014},
  \emph{89}, 165119\relax
\mciteBstWouldAddEndPuncttrue
\mciteSetBstMidEndSepPunct{\mcitedefaultmidpunct}
{\mcitedefaultendpunct}{\mcitedefaultseppunct}\relax
\EndOfBibitem
\bibitem[Fröhlich(1950)]{Frohlich1950}
Fröhlich,~H. Theory of the Superconducting State. I. The Ground State at the
  Absolute Zero of Temperature. \emph{Physical Review} \textbf{1950},
  \emph{79}, 845--856\relax
\mciteBstWouldAddEndPuncttrue
\mciteSetBstMidEndSepPunct{\mcitedefaultmidpunct}
{\mcitedefaultendpunct}{\mcitedefaultseppunct}\relax
\EndOfBibitem
\bibitem[Kruk and Kivshar(2022)Kruk, and Kivshar]{Kruk2022}
Kruk,~S.; Kivshar,~Y. Nonlinear photonics and topology: from fundamentals to
  applications. \emph{Light: Science \& Applications} \textbf{2022}, \emph{11},
  10\relax
\mciteBstWouldAddEndPuncttrue
\mciteSetBstMidEndSepPunct{\mcitedefaultmidpunct}
{\mcitedefaultendpunct}{\mcitedefaultseppunct}\relax
\EndOfBibitem
\bibitem[Sipe and Shkrebtii(2000)Sipe, and Shkrebtii]{SipeShkrebtii2000}
Sipe,~J.~E.; Shkrebtii,~A.~I. Second-order optical response in semiconductors.
  \emph{Phys. Rev. B} \textbf{2000}, \emph{61}, 5337--5352\relax
\mciteBstWouldAddEndPuncttrue
\mciteSetBstMidEndSepPunct{\mcitedefaultmidpunct}
{\mcitedefaultendpunct}{\mcitedefaultseppunct}\relax
\EndOfBibitem
\bibitem[Wismer \latin{et~al.}(2018)Wismer, Klamroth, and
  Saalfrank]{Wismer2018}
Wismer,~M.; Klamroth,~T.; Saalfrank,~P. Gauge-invariant formulation of
  light–matter interaction with ultrashort pulses. \emph{Phys. Rev. Lett.}
  \textbf{2018}, \emph{121}, 233902\relax
\mciteBstWouldAddEndPuncttrue
\mciteSetBstMidEndSepPunct{\mcitedefaultmidpunct}
{\mcitedefaultendpunct}{\mcitedefaultseppunct}\relax
\EndOfBibitem
\bibitem[Di~Ventra \latin{et~al.}(2002)Di~Ventra, Pantelides, and
  Lang]{Ventra2004}
Di~Ventra,~M.; Pantelides,~S.~T.; Lang,~N.~D. Current-induced forces in
  molecular wires. \emph{Phys. Rev. Lett.} \textbf{2002}, \emph{88},
  046801\relax
\mciteBstWouldAddEndPuncttrue
\mciteSetBstMidEndSepPunct{\mcitedefaultmidpunct}
{\mcitedefaultendpunct}{\mcitedefaultseppunct}\relax
\EndOfBibitem
\bibitem[Parker \latin{et~al.}(2019)Parker, Roy, and Das~Sarma]{Parker2019}
Parker,~D.~E.; Roy,~B.; Das~Sarma,~S. Topological quantum geometry in multiband
  systems. \emph{Phys. Rev. B} \textbf{2019}, \emph{100}, 041109(R)\relax
\mciteBstWouldAddEndPuncttrue
\mciteSetBstMidEndSepPunct{\mcitedefaultmidpunct}
{\mcitedefaultendpunct}{\mcitedefaultseppunct}\relax
\EndOfBibitem
\bibitem[Fregoso(2018)]{Fregoso2018}
Fregoso,~B.~M. Berry curvature and nonlinear optical response in time-reversal
  invariant insulators. \emph{Phys. Rev. B} \textbf{2018}, \emph{98},
  081110(R)\relax
\mciteBstWouldAddEndPuncttrue
\mciteSetBstMidEndSepPunct{\mcitedefaultmidpunct}
{\mcitedefaultendpunct}{\mcitedefaultseppunct}\relax
\EndOfBibitem
\bibitem[Schlawin \latin{et~al.}(2018)Schlawin, Tan, and Mukamel]{Schlawin2018}
Schlawin,~F.; Tan,~H.~T.; Mukamel,~S. Entangled two-photon absorption
  spectroscopy of excitons in semiconductors. \emph{Nature Communications}
  \textbf{2018}, \emph{9}, 2506\relax
\mciteBstWouldAddEndPuncttrue
\mciteSetBstMidEndSepPunct{\mcitedefaultmidpunct}
{\mcitedefaultendpunct}{\mcitedefaultseppunct}\relax
\EndOfBibitem
\bibitem[Rice and Mele(1982)Rice, and Mele]{RiceMele1982}
Rice,~M.~J.; Mele,~E.~J. Elementary Excitations of a Linearly Conjugated
  Diatomic Polymer. \emph{Physical Review Letters} \textbf{1982}, \emph{49},
  1455--1459\relax
\mciteBstWouldAddEndPuncttrue
\mciteSetBstMidEndSepPunct{\mcitedefaultmidpunct}
{\mcitedefaultendpunct}{\mcitedefaultseppunct}\relax
\EndOfBibitem
\bibitem[Zeng \latin{et~al.}(2021)Zeng, Liu, Jiang, Sun, and Xie]{Zeng2021}
Zeng,~J.; Liu,~H.; Jiang,~H.; Sun,~Q.-F.; Xie,~X.~C. Multiorbital model reveals
  a second-order topological insulator in 1H transition metal dichalcogenides.
  \emph{Phys. Rev. B} \textbf{2021}, \emph{104}, L161108\relax
\mciteBstWouldAddEndPuncttrue
\mciteSetBstMidEndSepPunct{\mcitedefaultmidpunct}
{\mcitedefaultendpunct}{\mcitedefaultseppunct}\relax
\EndOfBibitem
\bibitem[Haldane(1988)]{Haldane1988}
Haldane,~F. D.~M. Model for a Quantum Hall Effect without Landau Levels:
  Condensed-Matter Realization of the 'Parity Anomaly'. \emph{Physical Review
  Letters} \textbf{1988}, \emph{61}, 2015--2018\relax
\mciteBstWouldAddEndPuncttrue
\mciteSetBstMidEndSepPunct{\mcitedefaultmidpunct}
{\mcitedefaultendpunct}{\mcitedefaultseppunct}\relax
\EndOfBibitem
\bibitem[Bittner and Tyagi(2025)Bittner, and Tyagi]{bittner2025statistical}
Bittner,~E.~R.; Tyagi,~B. Statistical Control of Relaxation and Synchronization
  in Open Anyonic Systems. \emph{arXiv preprint arXiv:2504.02173}
  \textbf{2025}, \relax
\mciteBstWouldAddEndPunctfalse
\mciteSetBstMidEndSepPunct{\mcitedefaultmidpunct}
{}{\mcitedefaultseppunct}\relax
\EndOfBibitem
\bibitem[Jung and Kim(2022)Jung, and Kim]{Jung2022}
Jung,~J.; Kim,~Y.-H. Hidden breathing kagome topology in hexagonal transition
  metal dichalcogenides. \emph{Phys. Rev. B} \textbf{2022}, \emph{105},
  085138\relax
\mciteBstWouldAddEndPuncttrue
\mciteSetBstMidEndSepPunct{\mcitedefaultmidpunct}
{\mcitedefaultendpunct}{\mcitedefaultseppunct}\relax
\EndOfBibitem
\bibitem[Quir\'os-Cordero \latin{et~al.}(2025)Quir\'os-Cordero, Rojas-Gatjens,
  G\'omez-Dominguez, Li, Perini, Stingelin, Correa-Baena, Bittner, Kandada, and
  Silva-Acuna]{quiroscordero2025resolving}
Quir\'os-Cordero,~V.; Rojas-Gatjens,~E.; G\'omez-Dominguez,~M.; Li,~H.;
  Perini,~C. A.~R.; Stingelin,~N.; Correa-Baena,~J.-P.; Bittner,~E.~R.;
  Kandada,~A. R.~S.; Silva-Acuna,~C. Resolving exciton and polariton
  multi-particle correlations in an optical microcavity in the strong coupling
  regime. 2025; \url{https://arxiv.org/abs/2404.14744}\relax
\mciteBstWouldAddEndPuncttrue
\mciteSetBstMidEndSepPunct{\mcitedefaultmidpunct}
{\mcitedefaultendpunct}{\mcitedefaultseppunct}\relax
\EndOfBibitem
\bibitem[Tyagi \latin{et~al.}(2024)Tyagi, Li, Bittner, Silva-Acuña, and
  Piryatinski]{Tyagi2024JPCL}
Tyagi,~B.; Li,~H.; Bittner,~E.~R.; Silva-Acuña,~C.; Piryatinski,~A.
  Noise-Induced Quantum Synchronization and Entanglement in a Quantum Analogue
  of Huygens’ Clock. \emph{The Journal of Physical Chemistry Letters}
  \textbf{2024}, \emph{15}, 2520--2526\relax
\mciteBstWouldAddEndPuncttrue
\mciteSetBstMidEndSepPunct{\mcitedefaultmidpunct}
{\mcitedefaultendpunct}{\mcitedefaultseppunct}\relax
\EndOfBibitem
\bibitem[Bittner and Tyagi(2025)Bittner, and Tyagi]{Bittner2025JCP}
Bittner,~E.~R.; Tyagi,~B. Noise-Induced Synchronization in Coupled Quantum
  Oscillators. \emph{The Journal of Chemical Physics} \textbf{2025},
  \emph{162}, 034102\relax
\mciteBstWouldAddEndPuncttrue
\mciteSetBstMidEndSepPunct{\mcitedefaultmidpunct}
{\mcitedefaultendpunct}{\mcitedefaultseppunct}\relax
\EndOfBibitem
\end{mcitethebibliography}

\end{document}